\documentstyle[11pt,aaspp4,epsf]{article}
\newcommand{\re}{r_{\rm e}}
\newcommand{\rc}{r_{\rm c}}
\newcommand{\thg}{\theta_{\rm G}}
\newcommand{\vc}[1]{\mbox{\boldmath$\mathrm#1$}}
\newcommand{\kms}{\rm km\,s^{-1}}

\newcommand{\xup}[1]{{\setbox0=\hbox{$^{\rm #1}$}%
\hbox to\wd0{\hskip-0.05em plus1fil.\hskip0pt plus2fil}\llap{\box0}}}%
\newcommand{\xupd}{\xup{\circ}}
\newcommand{\xupa}{\xup{\prime\prime}}

\begin{document}
\title{High resolution optical and near-IR imaging of the quadruple
quasar RX~J0911.4+0551\altaffilmark{1}}
\altaffiltext{1}{Based on observations obtained at NOT and ESO La Silla}
\author{I. Burud}
\affil{Institut d'Astrophysique et de G{\'e}ophysique, Universit\'e de Li\`ege,\\
Avenue de Cointe 5, B--4000 Li\`ege, Belgium}

\author{F. Courbin}
\affil{Institut d'Astrophysique et de G{\'e}ophysique, Universit\'e de Li\`ege,\\
Avenue de Cointe 5, B--4000 Li\`ege, Belgium\\
URA 173 CNRS-DAEC, Observatoire de Paris,
F--92195 Meudon Principal C\'edex, France}

\author{C. Lidman}
\affil{ESO, Casilla 19001, Santiago 19, Chile}
 
\author{A.O. Jaunsen}
\affil{Institute of Theoretical Astrophysics, University of Oslo, Pb. 1029 Blindern, N-0315 Oslo, Norway\\
Centre for Advanced Study, Drammensvn. 78, N-0271 Oslo, Norway}

\author{J. Hjorth}
\affil{NORDITA, Blegdamsvej 17, DK-2100 Copenhagen {\O}, Denmark\\
Centre for Advanced Study, Drammensvn. 78, N-0271 Oslo, Norway}

\author{R. {\O}stensen}
{\affil{Institut d'Astrophysique et de G{\'e}ophysique, Universit\'e de Li\`ege,\\
Avenue de Cointe 5, B--4000 Li\`ege, Belgium\\
Department of Physics, University of Troms{\o}, N-9037 Troms{\o}, Norway}

\author{M.I. Andersen, J.W. Clasen}
{\affil{Nordic Optical Telescope, Apartado 474, E-38700 St. Cruz de La Palma,
Canary Islands, Spain}}

\author{O. Wucknitz}
\affil{Hamburger Sternwarte, Gojenbergsweg 112, D-21029 Hamburg, Germany} 

\author{G. Meylan}
\affil{ESO, Karl-Schwarzschild-Strasse 2, D-85748, Garching bei M\"unchen, Germany}

\author{P. Magain\altaffilmark{2}}
\affil{Institut d'Astrophysique et de G{\'e}ophysique, Universit\'e de Li\`ege,\\
Avenue de Cointe 5, B--4000 Li\`ege, Belgium}
\altaffiltext{2}{Ma\^{\i}tre de Recherches au 
Fonds National Belge de la Recherche Scientifique}
 
\author{R. Stabell}
\affil{Institute of Theoretical Astrophysics, University of Oslo,
Pb. 1029 Blindern, 0315 Oslo, Norway\\
Centre for Advanced Study, Drammensvn. 78, N-0271 Oslo, Norway,}

\author{S. Refsdal}
\affil{Hamburger Sternwarte, Gojenbergsweg 112, D-21029 Hamburg, Germany\\ 
Institute of Theoretical Astrophysics, University of Oslo,
Pb. 1029 Blindern, N-0315 Oslo, Norway\\
Centre for Advanced Study, Drammensvn. 78, N-0271 Oslo, Norway}

\begin{abstract}

We report   the detection of  four  images in the  recently discovered
lensed QSO RX~J0911.4+0551.   With   a maximum angular   separation of
$3.1$\arcsec, it  is the  quadruply imaged  QSO  with the widest known
angular separation.  Raw and deconvolved data reveal an elongated lens
galaxy.  The observed reddening in at least two of the four QSO images
suggests differential extinction by this lensing galaxy.  We show that
both an ellipticity of  the galaxy ($\epsilon_{\rm min}=0.075$) and an
external shear ($\gamma_{\rm min}=0.15$) from  a nearby mass has to be
included  in the lensing  potential in order  to reproduce the complex
geometry observed in  RX~J0911.4+0551.  A  possible galaxy cluster  is
detected about  38\arcsec\, from  RX~J0911.4+0551 and could contribute
to the X-ray emission  observed by ROSAT in  this field.  The color of
these galaxies indicates a plausible redshift in the range of 0.6-0.8.

\end{abstract}


\section{Introduction}

RX~J0911.4+0551, an AGN candidate  selected   from the ROSAT   All-Sky
Survey (RASS)  (Bade et al.  1995, Hagen  et  al.  1995), has recently
been classified  by Bade   et al.  (1997;    hereafter B97) as   a new
multiply imaged  QSO.   B97 show that  it consists  of  at least three
objects: two barely  resolved    components and a third   fainter  one
located 3.1\arcsec\ away from the other two.  They  also show that the
spectrum of this  third fainter component is  similar to the  combined
spectrum  of the two bright components.   The lensed source is a radio
quiet  QSO at $z=2.8$.   Since RASS detections  of distant radio quiet
QSOs are rare,   B97 pointed out   that the observed X-ray  flux might
originate from a galaxy cluster at $z \geq 0.5$ within the ROSAT error
box.  We present here  new optical and near-IR  high-resolution images
of RX~J0911.4+0551  obtained with the   2.56m Nordic Optical Telescope
(NOT) and   the  ESO  3.5m New  Technology   Telescope  (NTT). Careful
deconvolution of the data allows us to clearly resolve the object into
four QSO components  and a lensing galaxy.   In  addition, a candidate
galaxy cluster is detected in the vicinity of the four QSO images.  We
estimate   its redshift from the   photometric  analysis of its member
galaxies.

\section{Observations and reductions}

We first observed RX~J0911.4+0551 in  the $K$-band with IRAC~2b on the
ESO/MPI 2.2m  telescope on November 12,  1997.   In spite of  the poor
seeing   conditions ($\sim$ 1.3\arcsec),  preliminary deconvolution of
the data made it  possible  to suspect the  quadruple nature  of  this
object.  Much better optical observations were obtained at the NOT (La
Palma, Canary Islands,  Spain).  Three 300s exposures through  the $I$
filter, with a   seeing of $\sim  0\farcs8$ were  obtained with ALFOSC
under   photometric  conditions     on   November 16,    1997.   Under
non-photometric,    but     excellent   seeing   conditions     ($\sim
0\farcs5-0\farcs8$),  three    300s  $I$-band  exposures,   three 300s
$V$-band and five 600s $U$-band exposures were taken with HIRAC on the
night of December 3, 1997.  The pixel scales for  HIRAC and ALFOSC are
$0\farcs1071$ and   $0\farcs186$,  respectively.   RX~J0911.4+0551 was
also the first  gravitational lens to be   observed with the  new wide
field near-IR instrument SOFI, mounted on the ESO 3.5m NTT.  Excellent
$K$ and  $J$ images were taken on  December 15, 1997,  and January 19,
1998 respectively.  The 1024 $\times$ 1024  Rockwell detector was used
with a pixel scale of $0\farcs144$.

The optical data were  bias subtracted and flat-field  corrected using
sky-flats. Fringe-correction was  also  applied to the   $I$-band data
from  ALFOSC.   Sky subtraction was carried   out by fitting low-order
polynomial surfaces  to  selected  areas of  the  frames.  Cosmic  ray
removal was finally  performed on the  data.   The infrared  data were
processed as explained in Courbin,  Lidman \& Magain  (1998), but in a
much more  efficient  way for SOFI  than  for IRAC~2b data,  since the
array used with the former instrument is  cosmetically superior to the
array used with the latter.

\section{Deconvolution of RX~J0911.4+0551}
  
All  images were deconvolved using  the new deconvolution algorithm by
Magain, Courbin \& Sohy  (1998; hereafter MCS).   The sampling of  the
images  was improved in the  deconvolution  process, i.e. the  adopted
pixel size in the   deconvolved image is  half the  pixel size  of the
original frames. The final resolution  adopted in each band was chosen
according to the signal-to-noise (S/N)   ratio of the data, the  final
resolution improving with the S/N.

Our NOT  HIRAC data   were deconvolved  down  to  the best  resolution
achievable   with the   adopted    sampling  (2  pixels  FWHM,    i.e.
$0\farcs1$),  whereas   the NOT/ALFOSC  frames   were deconvolved to a
resolution of 3 pixels FWHM,  i.e. $0\farcs29$.  Although of very good
quality, the near IR-data have a lower S/N than the optical data.  The
resolution was therefore limited to 5 pixels FWHM in  both $J$ and $K$
($0\farcs36$) in order to avoid noise enhancement.

The MCS algorithm can be used  to deconvolve simultaneously a stack of
individual images   or to deconvolve  a  single  stacked  image.   The
results  from   the  simultaneous deconvolutions    are displayed   in
Fig.~\ref{fig:dec}.  The  four QSO images  are labeled A1, A2, A3, and
B.  The quality of the results was checked  from the residual maps, as
explained in  Courbin et  al.   (1998).   The deconvolution  procedure
decomposes the  images into a  number  of Gaussian  point sources, for
which  the program  returns  the  positions  and intensities, plus   a
deconvolved  numerical    background.   In   the   present   case, the
deconvolution was also performed   with an analytical De  Vaucouleurs,
and exponential disk galaxy  profile  at the  position of  the lensing
galaxy, in order to better describe its morphology.

\subsection{Results}

Table  1 lists the  flux of  each QSO component,  relative  to  A1, as
derived from the simultaneous deconvolutions.  Although the HIRAC $U$,
$V$ and  $I$ band data were taken  during  non photometric conditions,
they can still  be used to  determine the relative fluxes between  the
four images  of  the  QSO.  The   errors in the   relative fluxes  are
determined from  the  simultaneous deconvolutions   and represent  the
1-$\sigma$ standard deviation in the peak intensities.

When flux calibration was possible, magnitudes were calculated and the
results are displayed in Table 2, which also contains the positions of
the four QSO components  relative to A1.   The astrometric errors  are
derived by comparing the positions  of the components in the different
bands.

The  deconvolved numerical background  is used to determine the galaxy
position from the  first order moment   of the light  distribution.  A
reasonable estimate of the error on the lens position was derived from
moment measurements through  apertures of varying  size and on several
images  obtained  by   running deconvolutions   with different initial
conditions.   $I$, $J$, and  $K$ magnitudes  for  the galaxy were also
estimated from the deconvolved background image by aperture photometry
($\sim1.2$\arcsec\,  aperture in diameter).   The numerical  galaxy is
elongated in all three bands and the position angle  of its major axis
is $\theta_{G}\simeq 140\pm  5^{\circ}$.  In  the near-IR, the  galaxy
looks like it is composed  of a bright  sharp  nucleus plus a  diffuse
elongated   disk.  However we can     not exclude  that the   observed
elongation is due  to an unresolved  blend of two  or more intervening
objects.    None of our  two  analytical  profiles  fit  perfectly the
galaxy.   The  De   Vaucouleurs profile ($\rm  e=1-  b/a=0.31\pm0.07$,
$\theta_{G}=130\pm10^{\circ}$)  fits  slightly  better     than    the
exponential disk light distribution,  but still produces residual maps
with values as  large as 1.5-2 per pixel,  compared with a $\chi^2$ of
2.5-5 per pixel for the exponential disk  profile. The ellipticity and
the position angle derived this way are very  uncertain due to the low
S/N of the data.  Much deeper observations will be required to perform
precise surface  photometry of the   lens(es) and to  draw a  definite
conclusion about its (their) morphology.

\subsection{Field photometry}

In  order to  detect any  intervening galaxy  cluster  which might  be
involved in the overall lensing   potential  and contributing to   the
X-ray emission observed by ROSAT, we  performed $I$, $J$, and $K$ band
photometry on  all  the galaxies  in a 2.5\arcmin\,  field  around the
lensed QSO.  A   composite color image was  also  constructed from the
frames  taken through these 3  filters, in order to directly visualize
any group of galaxies with similar  colors, and therefore likely to be
at  the    same   redshift.  The  color  composite    is  presented in
Fig.~\ref{fig:field}.

Aperture   photometry was  carried  out using  the  SExtractor package
(LINUX version 1.2b5, Bertin  \& Arnouts, 1996). The  faintest objects
were   selected    to  have  at   least    5  adjacent   pixels  above
1.2$\sigma_{sky}$, leading to the limiting  magnitudes 23.8, 21.6, and
20.0  $\rm    mag/arcsec^{2}$  in   the  $I$,   $J$,   and $K$  bands,
respectively.  The faintest  extended object measured in the different
bands  had  magnitudes 23.0,  22.0, and   20.3 in  $I$, $J$, and  $K$,
respectively.      The  color-magnitude  diagram  of     the field was
constructed  from the $I$  and  $K$ band data  which  give  the widest
wavelength range possible with our photometric data.  Since the seeing
was different in the two  bands, particular attention  was paid to the
choice  of the isophotal apertures  fitted to the galaxies.  They were
chosen  to   be  as  large  as  possible,   still avoiding  too   much
contamination from the sky  noise, as an oversized  isophotal aperture
would introduce.   The color-magnitude diagram  of the galaxies in the
2.5\arcmin\, field is displayed in Fig.~\ref{fig:cmd} .  Stars are not
included in  this plot.  Note   that, because of their  proximity, the
magnitudes of the two blended members of the cluster candidate (center
of the  circle in Fig.~\ref{fig:field}) might  be underestimated by as
much as 0.3-0.4 magnitudes in both $I$ and $K$.

\section{Models}

In a first    attempt to model the   system  we chose  an   elliptical
potential of the form
\begin{eqnarray}
\psi &=& \psi (\re), \\
\re^2 &=& \frac{x'^2}{(1+\epsilon)^2} +
\frac{y'^2}{(1-\epsilon)^2},
\end{eqnarray}
where  the coordinates $x'$ and $y'$  are measured along the principal
axes  of the galaxy, whose position  angle $\thg$ is a free parameter.
For  small  ellipticities  $\epsilon$,  this    potential is   a  good
approximation  for  elliptical mass distributions  (Kassiola \& Kovner
1993). Additionally, an external  shear $\gamma$ with direction $\phi$
is included.  Even without  assuming an explicit potential  $\psi$, we
can determine the minimal $\gamma$ and  $\epsilon$ needed to reproduce
the observed image positions, by applying the methods given by Witt \&
Mao   (1997).  Their methods  eliminate  the unknown parameters of the
model to find constraints  for the ellipticity of  the galaxy  and the
external shear.  For  the shear we predict  a  minimum of $\gamma_{\rm
min}=0.15\pm0.07$, while $\epsilon_{\rm min}=0.075\pm0.034$ (both with
1$\sigma$ errors).  Therefore, neither ellipticity nor shear should be
omitted from  the    modeling.  To  keep  models simple,    we  used a
pseudo-isothermal  potential  corresponding to  an apparent deflection
angle of
\begin{equation}
\vc\alpha = \frac{\alpha_0}{\rc+\sqrt{\rc^2+\re^2}}
\,\left(\frac{x}{(1+\epsilon)^2}\,,\, \frac{y}{(1-\epsilon)^2} \right) ,
\end{equation}
which degenerates to a singular isothermal sphere for vanishing
core-radius $\rc$ and ellipticity $\epsilon$.

The model's parameters are determined by minimizing the $\chi^2$ given
by the observed and predicted positions of  the images and the galaxy.
The best model leads to $\chi^2 = 0.65$ and has parameters as given in
Table~\ref{tab:bestmod}. This value is reasonable for a model with one
degree of freedom ($\rc$ can not be  counted as a free parameter here,
because without the restriction  of $\rc\ge0$ it would become negative
in the   fit).    The positions  of the    images and  the   galaxy in
Table~\ref{tab:bestmod}  are parameters of the   model and have to  be
compared   with   the  observed  positions   in  Table~\ref{tab:phot}.

As shown above,    no elliptical potential  without a   shear,  and no
spherical potential with  shear can reproduce the  observational data.
For our pseudo-isothermal model, this results in large $\chi^2$ values
of  50.5 and 63.7, respectively.  These  values are much  too high for
the two degrees of freedom.

\section{Discussion}

Thanks    to   our   new high-resolution    imaging    data,   the QSO
RX~J0911.4+0551   is  resolved  into   four   images.  In    addition,
deconvolution with the  new MCS algorithm  reveals  the lensing galaxy,
clearly  confirming the   lensed nature  of  this system.    The image
deconvolution provides precise photometry and  astrometry for all  the
components of the system.

Reddening in components A2 and A3 relative to A1  is observed from our
$U$, $V$,  and $I$ frames  that were taken  within  three hours on the
same    night.  The  absence  of reddening    in  component  B and the
difference  in    reddening  between components    A2 and   A3 suggest
extinction by  the deflecting galaxy.  Note that although  our near-IR
data were obtained from  15 days to  6 weeks after the optical images,
they appear to be consistent with  the optical fluxes measured for the
QSO images, i.e.   flux ratios increase continuously  with wavelength,
from $U$ to $K$, indicating extinction by the lensing galaxy.

We have discovered a good galaxy cluster  candidate in the SW vicinity
of RX~J0911.4+0551 from our field photometry  in the $I$, $J$, and $K$
bands.  Comparison of our color-magnitude diagram with that of a blank
field (e.g., Moustakas et  al.  1997) shows  that the galaxies  around
RX~J0911.4+0551  are  redder  than  field-galaxies   at an  equivalent
apparent   magnitude.    In   addition,  the  brightest    galaxies in
Fig.~\ref{fig:cmd} lie on a red sequence at $I-K\sim 3.3$, typical for
the early  type members of  a distant galaxy  cluster.  The two dashed
lines indicate   our $\pm0.4$ color error   bars at $K\sim  19$ around
$I-K\sim3.3$. Most of these galaxies  are grouped in the region around
a double elliptical at a distance of  $\sim38$\arcsec\, and a position
angle of $\sim204^{\circ}$  relative to A1.   This can also be seen in
Fig.~\ref{fig:field} which shows a group  of red galaxies with similar
colors centered    on the double  elliptical  (in  the  center  of the
circle).

Consequently, there  is considerable evidence  for at least one galaxy
cluster in the field.  The redshift of our best candidate cluster (the
one circled in Fig.~\ref{fig:field}) can be estimated from the $I$ and
$K$ band photometry.  We have  compared the $K$-band magnitudes of the
brightest   cluster galaxies with  the    empirical $K$ magnitude  vs.
redshift  relation found by  Arag{\'o}n-Salamanca  et al.  (1998).  We
find that our cluster candidate,  with its brightest $K$ magnitude  of
about $\sim17.0$, should have  a  redshift of $z\sim0.7$.   A  similar
comparison  has been done in the  $I$-band without taking into account
galaxy  morphology.  We compare the  mean $I$ magnitude of the cluster
members with the  ones found by Koo et  al.  (1996)  for galaxies with
known redshifts in the Hubble Deep Field and obtain a cluster redshift
between  0.6 and 0.9.  Finally, comparison  of the  $I-K$ color of the
galaxy sequence   with data  and models from   Kodama  et  al.  (1998)
confirm the redshift estimate of 0.6-0.8.

In  order to calculate physical  quantities  from the model parameters
found in section 4, we assume a simple model for the cluster which may
be responsible  for the external shear.   For an isothermal potential,
the true shear and convergence are of the same order of magnitude.  As
the  convergence is not explicitly  included in the model, the deduced
shear is a reduced shear leading to an absolute convergence of $\kappa
=   \gamma/(1+\gamma)  = 0.241$.  For  a   cluster redshift of $z_{\rm
d}=0.7$ and  with cosmological parameters $\Omega=1$, $\lambda=0$ this
corresponds to   a velocity dispersion  of about   $1100\,\kms$ if the
cluster is  positioned at  an angular  distance  of 40\arcsec\,.   See
Gorenstein, Falco \& Shapiro (1988) for a discussion of the degeneracy
preventing a direct determination of $\kappa$.   From the direction of
the shear $\phi$,  (see  Table ~\ref{tab:bestmod}) we can  predict the
position angle of the cluster as seen from the QSO to be $12^\circ$ or
$192^\circ$.  The  latter value agrees well  with  the position of our
cluster candidate SW of the QSO images.   Note also the good agreement
between  the position angle $\thg$    derived from the observed  light
distribution,  and the predicted  position angle  corresponding to our
best fitting model of the lensing potential. Interestingly, this is in
good agreement  with Keeton,  Kochanek \& Falco  (1998) who  find that
projected mass distributions  are generally aligned with the projected
light distributions to less than $10^{\circ}$.

The color of the  main lensing galaxy is very  similar to that of  the
cluster members, suggesting that it might be a  member of the cluster.
Using the same model for the cluster  as above, assuming the galaxy at
the same redshift as the cluster, and neglecting the small ellipticity
of $\epsilon<0.05$, the velocity dispersion  of the lensing galaxy can
be predicted from the calculated deflection angle  $\alpha_0$ to be of
the  order of $240\,\kms$.  Since the  galaxy profile is sharp towards
the  nucleus in $K$, we   cannot rule out  the  possibility of a fifth
central image of  the   source,  as predicted for  non-singular   lens
models.   Near-IR    spectroscopy  is   needed  to    get  a  redshift
determination of the  lens and to  show  whether it is blended  or not
with a fifth image of the (QSO) source.

Some 10\arcsec\,  SW from the  lens, we detect a   small group of even
redder   objects.    These     red    galaxies  can   be      seen  in
Fig.~\ref{fig:field} a few arcseconds to the left  and to the right of
the cross. They  might be part  of a  second galaxy-group at  a higher
redshift, and   with  a position in  better   agreement with the X-ray
position  mentioned by B97.  However,  since the measured X-ray signal
is near the detection limit, and the 1-$\sigma$ positional uncertainty
is at least 20\arcsec\,, the  X-ray emission  is compatible with  both
the QSO  and  these galaxy  groups  in  the field.  Furthermore,  this
second group, at $z>0.7$, would most likely  be too faint in the X-ray
domain to be detected in  the RASS.  In fact,  even our lower redshift
cluster candidate would need to have  an X-ray luminosity of the order
of   $\rm  L_{0.1-2.4 \rm   keV}\sim  7.10^{44}  \rm erg\,\rm  s^{-1}$
(assuming    a   6    keV    thermal     spectrum,  $\rm    H_{0}=50\,
\rm{km\,s}^{-1}\,\rm Mpc^{-1}$,   $\rm q_{0}=0.5$),  in  order to   be
detected  with  0.02 $\rm cts\,\rm   s^{-1}$ by  ROSAT.  This is  very
bright but not  unrealistic for high  redshift galaxy clusters  (e.g.,
MS~1054-03, Donahue, Gioia, Luppino et al. 1997).

RX~J0911.4+0551 is  a new quadruply imaged QSO  with an  unusual image
configuration. The lens configuration is complex, composed of one main
lensing galaxy plus external shear possibly caused by a galaxy cluster
at redshift between 0.6 and 0.8 and another possible group at $z>0.7$.
Multi-object  spectroscopy is needed in  order  to confirm our cluster
candidate/s and derive  its/their redshift and velocity dispersion.
In addition, weak lensing analysis  of background galaxies might prove
useful to  map the overall  lensing potential involved in this complex
system.

\acknowledgements
 
The authors would like to thank  A. Moorwood and  J.G. Cuby for making
available to  us  the NTT/SOFI data  and the  referee  E.E.  Falco for
useful  comments.  We  also thank S.     Sohy  for her help   with the
preliminary data analysis, T.  Kodama  for providing his model results
for our  photometric system, T.J.  Broadhurst,  C.R.  Keeton, G.  Rauw
for helpful discussions and R. Sip for  help in improving the figures.
IB,  FC  and  R{\O} are supported    in part by  contract ARC94/99-178
``Action de  Recherche  Concert\'ee de  la Communaut\'e  Fran\c{c}aise
(Belgium)'' and P\^ole  d'Attraction Interuniversitaire, P4/05  (SSTC,
Belgium).

\begin{figure}
\begin{center}
\leavevmode  
\epsfxsize=11cm    
\caption{High resolution images of RX~J0911.4+0551 are displayed in
the  left   panels with, in  the    right panels, their  corresponding
deconvolved versions with  improved resolution and sampling.  In  each
band,  the  individual  data  frames  are  deconvolved simultaneously.
[Top:] Stack of 3 NOT+HIRAC I-band  image.  The total exposure time is
900s.  [Middle:]  Stack  of  9 NTT+SOFI   $J$-band frames.  The  total
exposure time is 1080s.  [Bottom:]  Stack of 19 NTT+SOFI $K$-band with
a total exposure time of 2400s.  For all the three bands the object is
clearly resolved into four  QSO images,  labeled A1,  A2, A3, and  B,
plus the elongated lensing galaxy.  The  field of the optical and near
IR data are  respectively 7\arcsec\, and  9\arcsec\, on a side.  North
is to the top and East to the left in all frames.}
\label{fig:dec}
\end{center}
\end{figure}

\begin{figure}
\begin{center}
\leavevmode
\epsfxsize=16cm
\caption{Composite    image     of  a    2\arcmin\,    field around
RX~J0911.4+0551. The frame has been obtained by combining our $I$, $J$
and $K$-band data. North   is up and East  to the left.  Note  the
group of red galaxies with similar colors, about 38\arcsec\, SW of the
quadruple lens (circle) and  the group of even redder  galaxies 10\arcsec\,
SW of the lens (cross).}
\label{fig:field}
\end{center}
\end{figure}

\begin{figure}
\epsfxsize=15cm  
\epsfbox{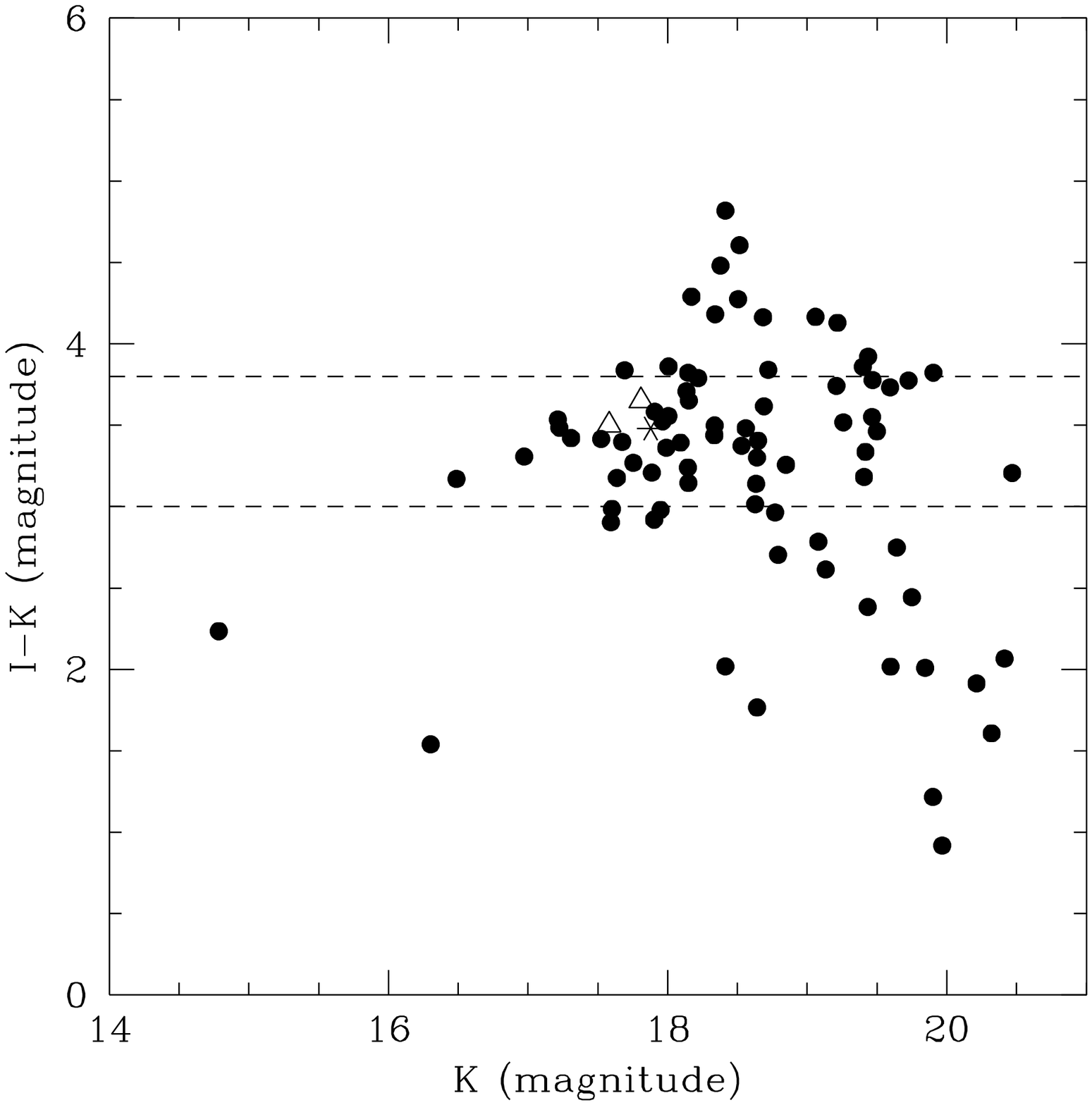} 
\caption{Color magnitude diagram
of the 2.5\arcmin\,  field around RX~J0911.4+0551.  The lensing galaxy
and many of    the  cluster  members  form   the   overdensity  around
$I-K\sim3.3$.  The lensing galaxy  in  RX~J0911.4+0551 is marked by  a
star and the  two   blended galaxies in  the  center  of  the  cluster
candidate  are plotted  as triangles.  Stars  are not  plotted in this
diagram.}
\label{fig:cmd}
\end{figure}

\begin{table}[t]
\begin{center}
\caption{Flux ratios for all four QSO components, relative to A1} 
\begin{tabular}{l r r r}
\hline \hline
    & A2  & A3  & B   \\
\hline
$F_{K}$ & $0.965\pm0.013$ & $0.544\pm0.025$  & $0.458\pm0.004$\\
$F_{J}$ & $0.885\pm0.003$ & $0.496\pm0.005$  & $0.412\pm0.005$\\
$F_{I}$ & $0.680\pm0.013$ & $0.398\pm0.002$  & $0.420\pm0.003$\\
$F_{V}$ & $0.587\pm0.009$ & $0.334\pm0.004$  & $0.413\pm0.006$   \\
$F_{U}$ & $0.590\pm0.013$ & $0.285\pm0.007$  & $0.393\pm0.004$    \\
\hline
\end{tabular}
\tablecomments{Results  obtained   from our non-photometric  data. All
measurements are given along with their 1-$\sigma$ errors.}
\end{center}
\label{tab:flux}
\end{table}

\begin{table}[t]
\begin{center}
\caption{Photometric and astrometric properties of RX~J0911.4+0551 
and the lensing galaxy.} 
\label{tab:phot}
\begin{tabular}{l r r r r r}
\hline \hline
    & A1  & A2  & A3  & B   & G \\
\hline
$K$ & $17.137\pm0.020$ & $17.308\pm0.035$ & $17.933\pm0.074$ & $18.119\pm0.010$ & $17.88\pm0.12$ \\
$J$ & $17.794\pm0.030$ & $17.927\pm0.029$ & $18.556\pm0.038$ & $18.757\pm0.020$ & $19.85\pm0.12$ \\
$I$  & $18.310\pm0.084$ & $18.728\pm0.096$ & $19.305\pm0.133$  &   $19.197\pm0.083$  & $21.36\pm0.13$ \\
$x$(\arcsec) & $0.000\pm0.004$ & $-0.259\pm0.007$ & $+0.013\pm0.008$ & $+2.935\pm0.002$ & $+0.709\pm0.026$ \\
$y$(\arcsec) &$0.000\pm0.008$ & $+0.402\pm0.006$ & $+0.946\pm0.008$ & $+0.785\pm0.003$ & $+0.507\pm0.046$ \\
\hline
\end{tabular}
\tablecomments{The astrometry is given  relative to component  A1 with
$x$ and  $y$  coordinates defined  positive to the   North and West of
A1. All measurements are given along with their 1-$\sigma$ errors. The
1-$\sigma$ errors on the photometric zero points are 0.02 in all bands.}
\end{center}
\end{table}

\begin{table}
\begin{center}
\caption{Parameters of the best-fit model.}
\label{tab:bestmod}
\begin{tabular}{cccccccc} \hline\hline
$\alpha_0$ & $\rc$ & $\epsilon$ & $\thg$ & $\gamma$ & $\phi$ & $\vc\theta_0$ &
$\vc\theta_{\rm S}$ \\ \hline
$1\xupa1061$ & $0$ & 0.0476 & $140\xupd5$ & 0.3176 &
$102\xupd3$ & $+0\xupa7078$ & $+1\xupa1459$ \\
&&&&&&$+0\xupa5440$ & $+0\xupa5452$ \\ \hline
\end{tabular}
\bigskip

\begin{tabular}{cccc} \hline\hline
image            & $x$ & $y$ & amplification \\ \hline
$\rm A_1$        & $+0\xupa0000$ & $-0\xupa0004$ & $-4.45$ \\
$\rm A_2$        & $-0\xupa2590$ & $+0\xupa4019$ & $+8.59$ \\
$\rm A_3$        & $+0\xupa0130$ & $+0\xupa9456$ & $-3.70$ \\
$\rm B$          & $+2\xupa9350$ & $+0\xupa7850$ & $+1.79$ \\ \hline
\end{tabular}
\tablecomments{$\vc\theta_0$ and $\vc\theta_{\rm S}$ are the position of the
  galaxy and the source, respectively.
The core radius $\rc$ was a free parameter constrained to $\rc\ge0$ in the
  modeling. The best fit lead to $\rc=0$.
Position angles $\thg$ and $\phi$ are measured from N to E. Negative
  amplification is due to odd parity of the image.}
\end{center}
\end{table}

\end{document}